\documentclass[lettersize,journal]{IEEEtran}
\usepackage{amsmath,amsfonts}
\usepackage{algorithm}
\usepackage{array}
\usepackage[caption=false,font=normalsize,labelfont=sf,textfont=sf]{subfig}
\usepackage{textcomp}
\usepackage{stfloats}
\usepackage{url}
\usepackage{verbatim}
\usepackage{graphicx}
\usepackage{cite}

\usepackage{amsthm,amsopn}
\usepackage{bm}
\usepackage{hyperref}
\usepackage{url}
\usepackage[capitalise]{cleveref}
\usepackage{graphicx,caption,lipsum}
\usepackage{booktabs,multirow}
\usepackage{placeins}
\usepackage{algpseudocode,algorithm}
\usepackage{color}
\usepackage[table,xcdraw]{xcolor}
\usepackage{bbm}
\usepackage{enumerate}   
\usepackage{adjustbox}
\usepackage[normalem]{ulem}
\usepackage{fixltx2e}
\usepackage{wrapfig}

\def\reg{{\rm\ooalign{\hfil
      \raise.07ex\hbox{\scriptsize R}\hfil\crcr\mathhexbox20D}}}
\usepackage{tikz}

\begin{document}

\title{Unsupervised TTS Acoustic Modeling for TTS with Conditional
Disentangled Sequential VAE}

\author{$^\ddagger$Jiachen Lian,~\IEEEmembership{Student Member,~IEEE}, $^\ddagger$Chunlei Zhang,~\IEEEmembership{Member,~IEEE}, Gopala K. Anumanchipalli, ~\IEEEmembership{Member,~IEEE}, Dong Yu,~\IEEEmembership{Fellow,~IEEE}
\thanks{Jiachen was an intern at Tencent AI Lab, Bellevue, WA. $^\ddagger$ Equal contribution. Chunlei Zhang is the corresponding author.}
}



\maketitle

\begin{abstract}
In this paper, we propose a novel unsupervised text-to-speech acoustic model training scheme, named UTTS, which does not require text-audio pairs. UTTS is a multi-speaker speech synthesizer that supports zero-shot voice cloning, it is developed from a perspective of disentangled speech representation learning. The framework offers a flexible choice of a speaker's duration model, timbre feature (identity) and content for TTS inference. We leverage recent advancements in self-supervised speech representation learning as well as speech synthesis front-end techniques for system development. Specifically, we employ our recently formulated Conditional Disentangled Sequential Variational Auto-encoder (C-DSVAE) as the backbone UTTS AM, which offers well-structured content representations given unsupervised alignment (UA) as condition during training. For UTTS inference, we utilize a lexicon to map input text to the phoneme sequence, which is expanded to the frame-level forced alignment (FA) with a speaker-dependent duration model. Then, we develop an alignment mapping module that converts FA to UA. Finally, the C-DSVAE, serving as the self-supervised TTS AM, takes the predicted UA and a target speaker embedding to generate the mel spectrogram, which is ultimately converted to waveform with a neural vocoder. We show how our method enables speech synthesis without using a paired TTS corpus in AM development stage. Experiments demonstrate that UTTS can synthesize speech of high naturalness and intelligibility measured by human and objective evaluations. Audio samples are available at our demo page~\footnote{\url{https://neurtts.github.io/utts\_demo/}}. 
\end{abstract}


\section{Introduction}
\IEEEPARstart{T}{ext-to-speech} (TTS) synthesis plays an important role for human computer interaction. With the continuous development of neural-based TTS systems, e.g., Tacotron~\cite{wang2017tacotron}, DurIAN~\cite{yu2020durian}, FastSpeech~\cite{ren2019fastspeech, fastspeech2, fastspeech2-multispeaker} or more recent Glow-TTS series~\cite{glowtts, tacotron2, vitsl,lim22_interspeech}, high-fidelity synthetic speech has been produced and the gap between machine generated speech and real speech is reduced. This is especially true for languages with rich resources, which refers to a sizeable high quality parallel speech and textual data. Usually, the supervised TTS system requires dozens of hours of single-speaker high quality data to retain a good performance. However, collecting and labeling such data is a non-trivial task, which is time-consuming and expensive. In that sense, current supervised solutions are still have their limitations on the demanding needs of ubiquitous deployment of customized speech synthesizers for AI assistants, gaming or entertainment industries. Natural, flexible and controllable TTS pathways become more essential when facing these diverse needs.

One possible way to relax the aforementioned limitation is to develop text-to-speech synthesis that does not require high quality parallel speech and textual data for training TTS acoustic models (AM). In this preliminary study, we propose a novel UTTS framework based on the recent advancement of self-supervised learning (SSL) techniques for speech representation~\cite{hsu2021hubert, wavlm,C-DSVAE-lian,DSVAE-VC}. Like many conventional supervised TTS systems, our proposed UTTS framework is essentially a two-stage TTS process, namely a model that converts a raw text to a mel spectrogram and a vocoder which transforms the mel spectrogram to the speech waveform. While the supervised TTS AM achieves the text to mel spectrogram transformation by explicitly composing an external aligner or by learning a latent alignment ~\cite{yu2020durian,fastspeech2,ren2019fastspeech,vitsl}, the UTTS AM must resolve this mapping with an indirect way since no paired TTS AM training data is available. 
\begin{wrapfigure}{rth}{5cm}
\centering
\includegraphics[width=0.27\textwidth]{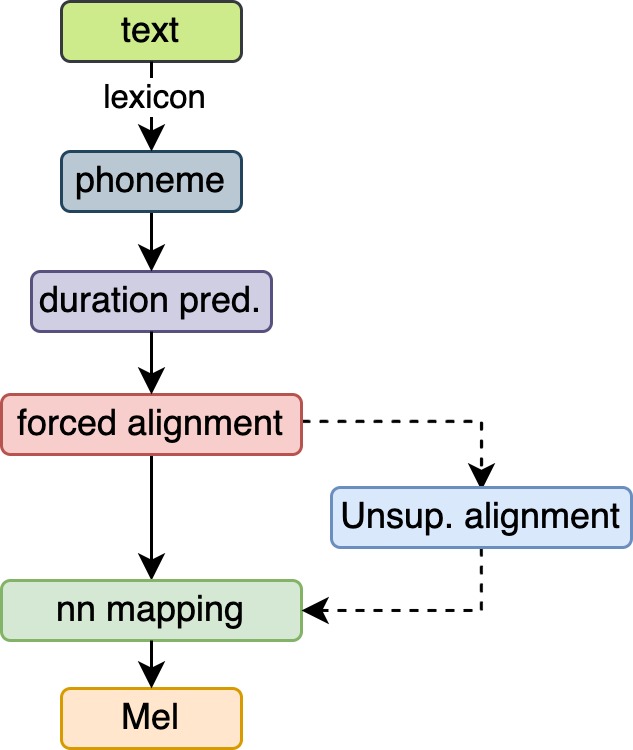}
\vspace{-2ex}
\caption{ Text to mel spectrogram converion with supervised TTS AM to UTTS AM }
\label{fs2utts}
\end{wrapfigure}

For conceptual comparison purpose, Fig.\ref{fs2utts} illustrates the necessary components of a typical supervised TTS AM (e.g., DurIAN~\cite{yu2020durian} or FastSpeech2~\cite{fastspeech2} which shares a similar structure) and our UTTS AM. Although UTTS follows most of the procedures of supervised AM during inference, the UTTS AM differs the supervised AM in three major ways: 1) For supervised AM, the neural network maps forced alignment of phoneme sequence to frame-by-frame acoustic features. We replace forced alignment with unsupervised alignment, which is derived from acoustic partitioning with the SSL features (e.g., kmeans labels of WavLM~\cite{wavlm} features). 2) supervised AMs (DurIAN and FastSpeech2) are only trained with TTS speech data, since UTTS investigates the Conditional Disentangled Sequential Variational Auto-encoder (C-DSVAE~\cite{C-DSVAE-lian}) as the backbone, the self-supervised/unsupervised model possibly enables more generic speech data to build a more generalized AM instead of being restricted with the TTS data only. 3) UTTS uses external speech recognition data to train the multi-speaker duration model and the projection between forced alignment to unsupervised alignment (FA2UA), while traditional TTS AMs rarely introduces such variability. 

For system implementation and experimental studies, the multi-speaker VCTK and LibriTTS  dataset~\cite{vctk2017,libritts} are utilized to demonstrate the effectiveness of the proposed method. We show the possibility that UTTS can generate speech of high naturalness and intelligibility in terms of objective and subjective tests. 

Although more detailed explanations and analysis can be found throughout this paper, the main contributions of this work are summarized as follows:
\begin{itemize}
    \item By exploring the conditions to end to end disentangled speech representation learning, we justify that unsupervised TTS AM is feasible.
    \item TTS and voice conversion are unified under a generative model (C-DSVAE), showing the potential of employing self-supervised (disentangled) representation learning for voice generation tasks.
    \item With disentangled representations, UTTS offers a flexible framework that can bridge control signal (speaker identity, text, style etc.) and generated voice, which enables more fine grained processes for controllable and customized TTS. 

\end{itemize}

The rest of this paper is organised as follows. Sec~\ref{related_word} introduces the background and related work. Section \ref{CDSVAE} illustrates the C-DSVAE and its development for voice generation. The UTTS framework is presented in Sec~\ref{Sec-UTTS}. Experimental setup and results are shown in Sec~\ref{exps}, with conclusions in Sec.~\ref{conclusion}.
\section{Related Work}
\label{related_word}
In this section, we introduce the background and relation to our proposed method.

\textbf{Supervised TTS:} the neural network based supervised TTS systems can be generally categorised into two forms, the end-to-end methods~\cite{oord2016wavenet, fastspeech2,glow_wavegan, vitsl, JETS} that directly maps text to speech waveform and the two-stage solutions~\cite{wang2017tacotron, yu2020durian, ren2019fastspeech, fastspeech2, fastspeech2-multispeaker, tacotron2,glow_wavegan} that utilize an acoustic model to convert text to frame by frame acoustic features (e.g., mel spectrogram) and a vocoder transforms the intermediate features to the waveform. While the end-to-end TTS draws an increased attention for some tasks, most of the current studies still follow the two-stage methods, mainly because of the less constraint from the data and optimization. As mentioned in the Introduction, the proposed UTTS system assembles like the two-stage supervised TTS, with the differences in some of the critical building blocks.  

\textbf{Unsupervised TTS:} there are not so many successful tries of the unsupervised TTS system development until recently. Thanks to the recent development of self-supervised speech representation learning, both systems in ~\cite{liu2022simple,ni2022unsupervised} used an unsupervised ASR model to transcribe the TTS speech data. With the pseudo labels, they once again can adopt the supervised TTS recipe to build the unsupervised TTS systems. The result indicates that it is possible to produce speech with high intelligibility even without paired TTS data. One notable concern for such unsupervised TTS systems remains in the development of unsupervised ASR, which still requires necessary improvement to close the gap between the supervised counterparts~\cite{baevski2021unsupervised,liu2022towards}. Strictly speaking, current unsupervised ASR systems have to rely on either lexicons or alignment prediction to match the acoustics and text, such lexicon or aligner are built with supervisions (e.g., paired data is used in developing CMU dictionary). Regarding our unsupervised approach, instead of transcribing the training speech data and performing supervised network mapping, our TTS AM is a variant of variational auto-encoder (i.e., C-DSVAE~\cite{C-DSVAE-lian}), with unsupervised alignment as the condition to regularize the learned latent representation to follow the phonetic structure. With this design, our UTTS system requires an additional mapping for text (forced alignment) space to the unsupervised alignment space. For that purpose, we just conduct a data driven process to map forced alignment to unsupervised alignment with the employment of an ASR system. Please refer to Section~\ref{fa2us} for more details about transforming text to unsupervised acoustic alignment for UTTS inference.      

\textbf{Disentangled speech representation for voice generation:} speech disentanglement algorithms, such as DSVAE and C-DSVAE ~\cite{hsu2017unsupervised,dsvae,DSVAE-VC,C-DSVAE-lian}, provide a way to isolate global timbre/speaker feature and local content features. Our work is inspired from disentangled speech representation learning and its downstream tasks, such as reconstruction/resynthesis and voice conversion~\cite{DSVAE-VC,C-DSVAE-lian,polyak2021speech,qian2019autovc}. The difference between those studies and UTTS is that, resyntheis or VC are speech driven, the process keeps the content unchanged, while UTTS has to make a linkage between raw text and the intermediate content embeddings in order to generate meaningful output speech. And how to connect the front-end text with unsupervised TTS AM (C-DSVAE) remains the key in building our UTTS system.     
\section{C-DSVAE for Voice Generation}
\label{CDSVAE}
\begin{figure*}[h]
    \centering
    \includegraphics[width=0.9\linewidth, height=7.2cm]{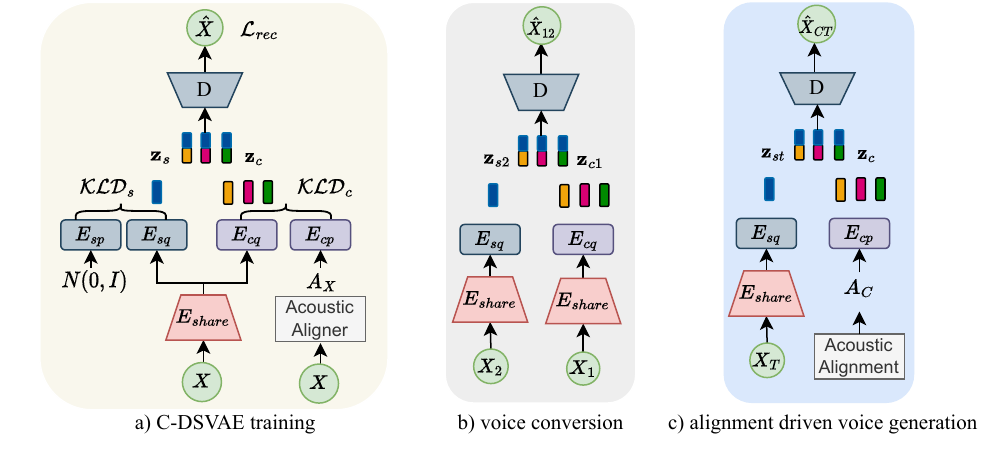}
    \vspace{-1ex}
    \caption{The system diagram of conditional DSVAE.}
    \label{dsvae fig}
\end{figure*}

\subsection{Overview}
Disentangling speaker information from speech content is of great importance to speech tasks such as speaker recognition, automatic speech recognition, voice conversion and text-to-speech~\cite{zhang2022c3,nagrani2020disentangled,autovc,jia2018transfer,wegmann1996speaker}. Of the existed speech disentanglement methods, DSVAE \cite{dsvae} is a compelling framework that generates the disentangled speaker and content representation in an unsupervised manner. A recent DSVAE study in speech \cite{DSVAE-VC} presents impressive speaker verification and voice conversion performance by balancing the information flow between speaker and content. As the content prior being assumed independent of the input data, early versions of DSVAE suffer from discontinuous and unstable speech generation with damaged phonetic structures \cite{s3vae, rwae, contras-dsvae,dsvae}. More recently, C-DSVAE \cite{C-DSVAE-lian} was proposed to introduce the content-dependent acoustic alignment as a condition to regularize the DSVAE training, from which the phonetically discriminative content representations are learned and the stabilized vocalization is achieved, overcoming the architecture drawback of earlier explorations in DSVAE \cite{s3vae, rwae, contras-dsvae} and delivering enhanced zero-shot VC performances. In the following sections, we first present a systematic overview of the DSVAE framework, which serves as the backbone of the improved C-DSVAE. Then, we present detailed development methods of C-DSVAE by deliberating different types of content conditions as well as training objectives. Furthermore, we extend C-DSVAE by introducing the \textit{masked prediction training} to capture the robust and long-range temporal relations over acoustic units. In the last two sections, we discuss how the voice conversion and alignment-driven voice generation are performed within the C-DSVAE framework, the latter exploration serves as the cornerstone of our UTTS system to be discussed in Sec.~\ref{Sec-UTTS}. The detailed model architectures can be found in Sec.~\ref{UTTS-model}.

\subsection{DSVAE Framework} \label{DSVAE-framework}
\subsubsection{Backbone}
Fig.~\ref{dsvae fig} (a) illustrates the training process of C-DSVAE, where DSVAE serves as the backbone
architecture. The DSVAE consists of a shared encoder $E_{share}$, a posterior speaker encoder $E_{sq}$, a posterior content encoder $E_{cq}$, a prior speaker encoder $E_{sp}$, a prior content encoder $E_{cp}$ and a decoder $D$. The input utterance $X$ is passed into $E_{share}$, followed by $E_{sq}$ and $E_{cq}$ which encode the speaker posterior distribution $q(z_s|X)$ and the content posterior distribution $q(z_c|X)$, respectively. $z_s$ denotes the speaker embedding and $z_c$ denotes the content embedding. For the prior modeling, $E_{sp}$ encodes the speaker prior $p(z_s)$ and $E_{cp}$ encodes the content prior $p(z_c)$. Depending on different voice generation tasks, during the decoding/generation stage, the speaker embedding $z_s$ and content embedding $z_c$ are sampled from either the posteriors $q(z_s|X)$ and $q(z_c|X)$ or the priors $p(z_s)$ and $p(z_c)$, and the concatenation of them is passed into decoder $D$ to generate the synthesized speech $\hat{X}=D(z_s, z_c)$.

\subsubsection{Probabilistic Graphical Models and Training Objective}
As introduced in \cite{dsvae, DSVAE-VC, C-DSVAE-lian}, DSVAE is designed to follow independence assumption for both priors and posteriors to implicitly leverage disentanglement between speaker and content representations. The probabilistic graphical models are formulated in Eq.~\ref{prior-independence} and Eq.~\ref{posterior-independence}. Note that the factorization of $p_{\theta}(z_c)$ and $q_{\theta}(z_c|X)$ can either follow the streaming manner \cite{dsvae, contras-dsvae, DSVAE-VC} or non-streaming manner \cite{C-DSVAE-lian}, depending on the practical implementation. Eq.~\ref{prior-independence} and Eq.~\ref{posterior-independence} correspond to the latter one, where $t$ is the frame index and $\theta$ denotes model parameters.
 \vspace{-1ex}
\begin{equation} \label{prior-independence} 
    p_{\theta}(z_s,z_c)=p(z_s)p_{\theta}(z_c)=p(z_s)\prod_{t=1}^T p_{\theta}(z_{ct}|z_{c})
\end{equation}

\begin{equation} \label{posterior-independence}
\begin{split}
     q_{\theta}(z_s,z_c|X) & =q_{\theta}(z_s|X)q_{\theta}(z_c|X) \\& =q_{\theta}(z_s|X)\prod_{t=1}^T q_{\theta}(z_{ct}|X)
     \end{split}
\end{equation}
The overall training objective for DSVAE is a VAE loss, but with two latent variables i.e. $z_s$ and $z_c$, as shown in Eq.~\ref{dsvaeloss}. $L_{DSVAE}$ forces strong disentanglement between $z_s$ and $z_c$ without additional constraints such as mutual information regularization \cite{s3vae, contras-dsvae, rwae, idevc}, as explained in \cite{DSVAE-VC}. KL divergence is denoted as $KLD$, $p_{\theta}(X|z_s,z_c))$ is modeled by the decoder $D$, representing the generation process. $\alpha$ and $\beta$ are two balancing factors which are used to control the disentanglement~\cite{DSVAE-VC}.
\begin{equation}\label{kls-dsvae}
    \mathcal{L}_\mathit{KLD_s}=\mathbb{E}_{p(X)}[ \mathit{KLD}(q_{\theta}(z_s|X)||p(z_s))]
\end{equation}

\begin{equation}\label{klc-dsvae}
    \mathcal{L}_\mathit{KLD_c}=\mathbb{E}_{p(X)}[\mathit{KLD}(q_{\theta}(z_c|X)||p_{\theta}(z_c))]
\end{equation}

\begin{equation}\label{dsvaeloss}
\begin{split}
    \mathcal{L}_\mathit{DSVAE}= & \mathbb{E}_{p(X)}\mathbb{E}_{q_{\theta}(z_{s}, z_{c}|X)}[-\log(p_{\theta}(X|z_s,z_c))] \\
     & +\alpha \mathcal{L}_\mathit{KLD_s}+\beta \mathcal{L}_\mathit{KLD_c}
    \end{split}
\end{equation}
\subsection{C-DSVAE System}
\label{conditional-prior}
In the DSVAE systems \cite{dsvae, s3vae, contras-dsvae, DSVAE-VC}, the content prior $p(z_c)$ is independent of $X$. As formulated in Eq.\ref{klc-dsvae}, the training objective is to minimize the KL divergence between $p(z_c)$ and $q(z_c|X)$. Since $q(z_c|X)$ explicitly models the phonetic structure of the speech content, the content-independent prior will force the content representation to lose phonetic discriminativity in the latent space, which is detrimental to the disentanglement. In order to generate phonetically meaningful and continuous speech
with stable vocalizations, we introduce the acoustic alignment $A_X$ as the condition for content prior distribution. The motivations are two-folds. First, acoustic alignment carries the phonetic structure of the utterance, which will be encoded into the content prior model. Such alignment-aware prior acts as a ``teacher" model to force the ``student", which is the content posterior $q_{\theta}(z_c|X)$, to inherit the phonetic structure of the original utterance. Second, during the inference, the acoustic alignment $A_X$ could be sampled from the alignment-aware prior so as to perform alignment-driven speech synthesis, which is the basis for our UTTS system to be introduced in Sec.~\ref{Sec-UTTS}. 
\subsubsection{Acoustic Alignment as Content Condition} We consider two types of acoustic alignment: forced alignment (\textbf{FA}) and unsupervised alignment (\textbf{UA}). For a specific utterance $X$, we denote the acoustic alignment as $A_{X}^{FA}$ and $A_{X}^{UA}$ respectively.

\textbf{Forced Alignment}: $A_{X}^{FA}$ is the forced alignment of the utterance $X$. We use \textit{Montreal forced alignment} (MFA) \cite{mfa} to extract the forced alignment given audio-text pair. In consistency with \cite{fastspeech2-multispeaker}, the forced alignment is a sequence of monophones and there is a total number of 72 monophones. In the next phase, the content prior encoder $E_{cp}$ takes the one-hot form of alignment sequence as input at each time step and predicts the frame-wise content prior distribution $p(z_c|A_{X}^{FA})$. 

\textbf{Unsupervised Alignment}: The forced alignment $A_X^{FA}$ is not always available since audio-transcription pair does not naturally exist. And because of this, the existing supervised TTS models are not scaled-well with respect to audio samples under diverse acoustic conditions, which limits their potential in synthesizing natural and expressive speech in the generalized open domains. To further utilize the advantage of untranscribed audio samples that are largely existed, we switch to the unsupervised alignment that is derived via the self-supervised pre-training. Of the popularly proposed self-supervised speech pre-training models, WavLM \cite{wavlm} delivers state-of-art performance in ``full stack" downstream speech tasks such as speaker recognition, speaker diarization, ASR and speech separation. WavLM employs a denoising masked speech prediction training strategy by feeding noisy and overlapped speech as inputs to predict the pseudo-labels of original speech on masked region, which improves both the speech content modeling capability and model robustness. The WavLM-Base model is pretrained from Librispeech-960h \cite{panayotov2015librispeech}, and we adopt the WavLM-Base \cite{wavlm} model to extract the acoustic features. With WavLM features adopted, Kmeans++ \cite{kmeans++} algorithm is performed on the features to generate frame-wise pseudo labels, denoting as unsupervised alignment $A_{X}^{UA}$, which is also assumed to capture the phonetic structure of the utterance $X$. The pretrained WavLM-Base model together with Kmeans algorithm acts as the \textit{acoustic aligner} as indicated in Fig.~\ref{dsvae fig} (a).

\subsubsection{Masked Unit Prediction Training} \label{masked-unit-prediction}
In order to capture the robust and long-range temporal relations over acoustic units and to generate more continuous speech, we adopt the \textit{Masked Unit Prediction} (MUP) when training the content prior encoder $E_{cp}$. Denote $M(A_X)\subset [T]$ as the collection of $T$ masked indices for a specific condition $A_X$, where the masking configuration is consistent with \cite{hsu2021hubert}. Let $\tilde{A_X}$ be a corrupted version of $A_X$, in which $A_{X_t}$ will be masked out if $t\in M(A_X)$. Denote $z_{cp}$ as the sample of output of the content prior encoder $E_{cp}$, i.e. $z_{cp}\sim E_{cp}(A_X)$. The negative loglikelihood loss (NLL) $\mathcal{L}_\mathit{MUP-C}$ for condition modeling is defined in Eq.~\ref{L-MUP-C}, where $p(z_{{cp}_i}|\tilde{A_{X_i}})$ is the softmax categorical distribution. $\mathbb{E}_{A_X}$ denotes the expectation over all $A_X$. Since it is reported in \cite{hsu2021hubert} that masked-only prediction delivers impressive ASR performances, we do not incorporate the prediction loss for unmasked regions. 
\begin{equation} \label{L-MUP-C}
    \mathcal{L}_\mathit{MUP-C}=-\mathbb{E}_{A_X}\Sigma_{i\in M(A_X)} \log p(z_{{cp}_i}|\tilde{A_{X_i}}) 
\end{equation}
\subsubsection{Training Objective and Configuration}
\label{training_config}
We adopt the DSVAE training objective as expressed in Eq.~\ref{dsvaeloss}, with the exception that the $L_{KLD_c}$ is formulated with conditional content prior, as shown in Eq.~\ref{klc-c-dsvae}. We design the content prior encoder $E_{cp}$ such that $p_{\theta}(z_c|A_X)$ follows a non-streaming factorization. The masked unit prediction loss $L_{MUP-C}$ is also incorporated. The overall loss objective for C-DSVAE is expressed in Eq.~\ref{cdsvaeloss}. When $\gamma=0$, it is vanilla C-DSVAE training. We set $\alpha=0.01$, $\beta=10$ and $\gamma=1$ if masked unit prediction training is involved. We adopt Adam optimizer~\cite{adam} with an initial learning rate of 5e-4 which is decayed by 0.95 every 5 epochs. The batch size is fixed to be 256. The training configurations are in line with~\cite{DSVAE-VC, C-DSVAE-lian}.
\begin{equation}\label{klc-c-dsvae}
\mathcal{L}_\mathit{KLD_c-C}=\mathbb{E}_{p(X)}[\mathit{KLD}(q_{\theta}(z_c|X)||p_{\theta}(z_c|A_X))]
\end{equation}
\begin{equation}\label{cdsvaeloss}
\begin{split}
\mathcal{L}_\mathit{C-DSVAE}= & \mathbb{E}_{p(X)}\mathbb{E}_{q_{\theta}(z_{s},z_{c}|X)}[-log(p_{\theta}(X|z_s,z_c))] \\ & +\alpha \mathcal{L}_\mathit{KLD_s}+\beta \mathcal{L}_\mathit{KLD_c-C}+\gamma L_\mathit{MUP-C}
\end{split}
\end{equation}
\subsection{Applications of C-DSVAE}\label{voice-conversion-overview}
\textbf{Voice Conversion}: VC seeks to convert the speaker information of source utterance to the target speaker while keeping the speech content unchanged. As presented in Fig.~\ref{dsvae fig} (b), given the arbitrary source utterance $X_1$ and arbitrary target utterance $X_2$, the speaker embedding $z_{s2}$ is extracted from $X_2$ using the posterior speaker encoder $E_{sq}$ and the content embedding $z_{c1}$ is extracted from $X_1$ using the posterior content encoder $E_{cq}$. The decoder then takes the concatenation of $z_{s2}$ and $z_{c1}$ and generates the converted mel spectrogram $\hat{X_{12}}$, which is converted to waveform via a pre-trained neural vocoder.

\textbf{Alignment driven Voice Generation}: Given acoustic alignment $A_C$ (either forced alignment or unsupervised alignment) and arbitrary target utterance $X_T$, alignment-driven voice generation aims to synthesize the speech from the alignment with the target speaker embedding, as shown in Fig.~\ref{dsvae fig} (c). Specifically, the speaker embedding $z_s$ is extracted from $X_T$ using posterior speaker embedding $E_{sq}$ and the content embedding is sampled from $p_{\theta}(z_c|A_C)$ generated via the content prior encoder $E_{cp}$, which takes $A_C$ as input. The alignment-driven voice generation is an essential trick that bridge the textual condition to the acoustic representations in our study, which serves as a cornerstone of our UTTS framework.

\section{UTTS System} \label{Sec-UTTS}

\begin{figure*}[ht]
    \centering
    \includegraphics[width=1\linewidth, height=6.8cm]{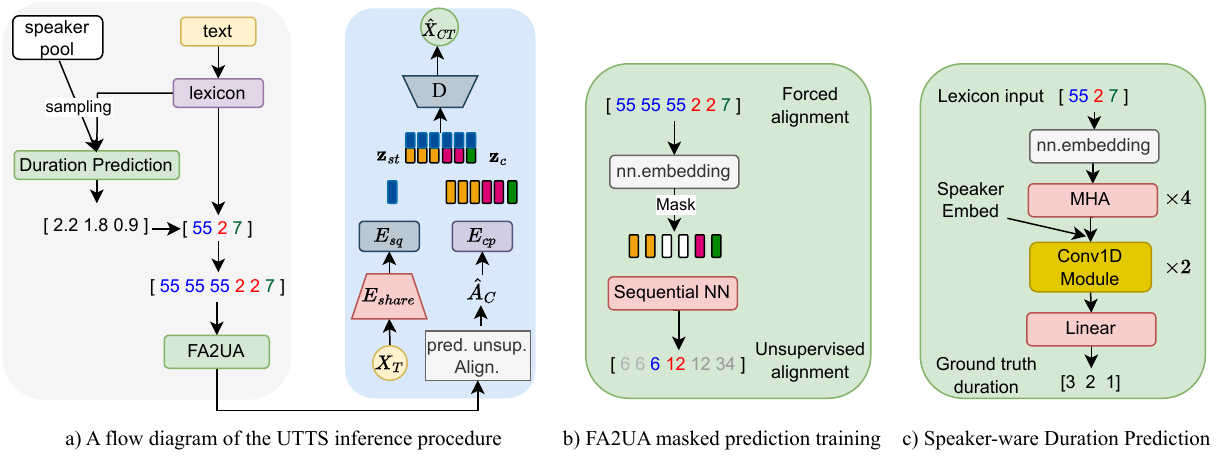}
    \vspace{-2ex}
    \caption{The core components of the proposed UTTS system.}
    \label{UTTS}
\end{figure*}

Our proposed UTTS pipeline is presented in Fig.~\ref{UTTS}. We first obtain the phoneme sequence of the text transcription with Librispeech Lexicon \footnote{\url{https://www.openslr.org/11/}}, which is consistent with \cite{fastspeech2-multispeaker}. The phoneme sequence is then converted to a list of token indices. The phoneme sequence (token indices) together with a speaker embedding (from a speaker pool for duration model training) is passed into the \textbf{Speaker-Aware Duration Prediction} module which delivers the predicted forced alignment (FA). The \textbf{FA2UA} module takes FA as input and predicts unsuperviserd alignment (UA). The UA along with a target speaker embedding is fed into the C-DSVAE to generate mel spectrogram. A neural vocoder is then applied to convert mel spectrogram to waveform. It is noted that the proposed UTTS system performs zero-shot \textbf{voice cloning} for the target utterance, which is a very difficult setting for TTS. In the following sections, we present the \textbf{Speaker-Aware Duration Prediction} and \textbf{FA2UA} modules. Both the modules are trained separately, the detailed model architecture of each module are presented in Sec.~\ref{UTTS-model}.

\subsection{Speaker-Aware Duration Prediction}
The duration predictor takes the phoneme sequence as well as speaker embedding as input to predict the speaker-aware duration for each phoneme~\cite{yu2020durian,fastspeech2}, as shown in Fig.~\ref{UTTS} (c). Specifically, the phoneme sequence is first passed into a trainable look-up table to obtain the phoneme embeddings. Afterwards, a four-layer multi-head attention (MHA) module is followed to extract the latent phoneme representation. A two-layer conv-1D module is then used to take the summation of latent phoneme representation and speaker embedding sampled from the speaker pool. A linear layer is finally applied to generate the predicted duration in the logarithmic domain. During training, the Mean Squared Error (MSE) is utilized to calculate the difference between the predicted duration and the target duration. For training, the target duration of a phoneme sequence is obtained from the forced alignment extracted by Montreal Forced Alignment (MFA)~\cite{mfa}. Note that the target duration is also in the logarithmic scale. During inference, we round up the duration generated by the predictor and expand the phoneme sequence to form an estimated forced alignment. We also present an example of how the duration predictor works during inference. Assume a random speaker is sample and the phoneme sequence is [55 2 7]. The predicted duration sequence is [2.2 1.8 0.9], correspondingly. We first round up the duration to make it become [3 2 1]. The phoneme sequence is then expanded to [55 55 55 2 2 7] as the predicted forced alignment (FA). The duration predictor works similarly to \cite{yu2020durian,fastspeech2,fastspeech2-multispeaker}.

\subsection{FA2UA with Masked Prediction Training}
\label{fa2us}
The FA2UA module takes forced alignment (FA) as input and predicts corresponding unsupervised alignment (UA). Specifically, FA is first passed into a learnable look-up table to obtain the FA embeddings. Subsequently, a 3-layer Bi-LSTM module is employed to predict the UA embeddings given FA embeddings. During training, we adopt a masked prediction training strategy to train our FA2UA module as masked prediction is expected to be good at capture the long-range time dependency across tokens and encode more contextualized information for each token. Similar as Sec.~\ref{masked-unit-prediction}, we denote $M(\mathit{FA})\subset [T]$ as the collection of $T$ masked indices for a specific unsupervised alignment $\mathit{FA}$. Let $\tilde{FA}$ be a corrupted version of $\mathit{FA}$, in which $\mathit{FA}_{i}$ will be masked out if $i\in M(\mathit{FA})$.  Denote ${\mathit{UA}}_i$ as the corresponding frame level unsupervised alignment of ${\mathit{FA}}_i$, which serves as the ground truth for the masked prediction training. The negative loglikelihood loss (NLL) $\mathcal{L}_{\mathit{FA2UA}}$ for masked prediction training is defined in Eq.~\ref{L-FA2UA}, where $p({\mathit{UA}}_i|\tilde{{\mathit{FA}}_{i}})$ is the softmax categorical distribution. $\mathbb{E}_\mathit{(FA,UA)}$ denotes the expectation over all $\mathit{(FA,UA)}$ pairs. Following Sec.~\ref{masked-unit-prediction}, masked-only prediction loss is incorporated for unsupervised alignment prediction. Fig.~\ref{UTTS} (b) presents an example, where [55 2] are masked out from the input FA and the model only predicts the corresponding UA targets [6 12] in the training phase. During the inference, we pick the token with the maximum probability $p({\mathit{UA}}_i|{\mathit{FA}}_i) $ at each time step $i$ to form the predicted UA sequence. 
\begin{equation} \label{L-FA2UA}
    \mathcal{L}_{\mathit{FA2UA}}=-\mathbb{E}_{\mathit{(FA,UA)}}\Sigma_{i\in M(\mathit{FA})} \log p({\mathit{UA}}_i|\tilde{{\mathit{FA}}_{i}}) 
\end{equation}

\subsection{Dual Content Encoders for Reconstruction}
It is noted that the content representation to the decoder of C-DSVAE is sampled from the posterior content encoder $E_{cq}$ in the original C-DSVAE training, which is suitable for the voice conversion tasks but not necessarily ideal for alignment-driven voice generation (i.e., UTTS inference). There exists a mismatch between the reconstruction based training and the UTTS inference. To alleviate this mismatch, we propose to use \textbf{dual content encoders for reconstruction} in the C-DSVAE training. Specifically, we run the normal C-DSVAE training guided by Eq.~\ref{cdsvaeloss} until convergence, and then we start a second round of training where we also sample latent representation $z_{cp}$ from the prior content encoder for reconstruction. And the second round of training is initialed with the pretrained C-DSVAE model. With this kind of training strategy, the dual content encoders for reconstruction training can be described as Eq.~\ref{cdsvaeloss2}:
\begin{equation}\label{cdsvaeloss2}
\begin{split}
\mathcal{L}_\mathit{C-DSVAE}= & \frac{1}{2}\{\mathbb{E}_{p(X)}\mathbb{E}_{q_{\theta}(z_{s},z_{cq}|X)}[-log(p_{\theta}(X|z_s,z_{cq}))] + \\ & \mathbb{E}_{p(X)}\mathbb{E}_{q_{\theta}(z_{s},z_{cp}|X,A_X)}[-log(p_{\theta}(X|z_s,z_{cp}))] \}  \\ & +\alpha \mathcal{L}_\mathit{KLD_s}+\beta \mathcal{L}_\mathit{KLD_c-C}+\gamma L_\mathit{MUP-C}
\end{split}
\end{equation}
where the new reconstruction term is formulated as an average loss with both posterior content embedding and prior content embedding for reconstruction. In our experiment, we find a clear quality improvement of the UTTS samples over the posterior content only reconstruction models.   
\section{Experiments}
\label{exps}
\subsection{datasets} \label{datasets}
\paragraph{VCTK} 
For VCTK corpus~\cite{vctk2017}, we exclude speaker p315 since the transcriptions are not available. we then randomly select 90\% of the speakers as training set and the remaining 10\% as test set. The data is downsampled to 16 KHz. We extract mel spectrogram with the window size/hop size of 64 ms/ 16 ms, and the feature dimension is 80. 36900 trials are randomly generated for speaker verification subtasks. The settings are consistent with~\cite{DSVAE-VC, C-DSVAE-lian}. 
\paragraph{LibriTTS}
We downsample the original LibriTTS~\cite{libritts} data to 16 kHz. The original training split is taken as training set for C-DSVAE, FA2UA and duration predictor training. The mel spectrogram that is extracted for training follows the same configurations as the VCTK dataset. 

\subsection{Implementation details}
\subsubsection{Model Architectures}
\label{UTTS-model}
Table.~\ref{UTTS-model-table} presents details of UTTS system. In the C-DSVAE architecture, for posterior speaker encoder $E_{sq}$, posterior content encoder $E_{cq}$ and prior content $E_{cp}$, there is one dense layer used to predict the mean and one other dense layer used to predict the standard deviation. A linear classifier will be added at the end of $E_{cp}$ if masked prediction training is involved. Both speaker and content dimension are set to 64. The decoder $D$ consists of a prenet $D_{pre}$ and a postnet $D_{post}$, which were introduced in~\cite{autovc, DSVAE-VC, C-DSVAE-lian}. We use HiFi-GAN~\cite{hifigan, libritts} pretrained with LibriTTS as the universial vocoder\footnote{\url{https://github.com/ming024/FastSpeech2/tree/master/hifigan}} for voice conversion and text-to-speech synthesis. We use the same mask configuration as~\cite{hsu2021hubert} for all mask prediction experiments. 

\begin{table*}[ht] 
\centering
    \begin{tabular}{|c|c|c|}
    \hline
    \multicolumn{3}{|c|}{UTTS Model Architecture} \\
    \hline
    \hline
    \multirow{8}{*}{C-DSVAE} & $E_{Share}$ & (Conv1D(256, 5, 2, 1)$\rightarrow$ InstanceNorm2D$\rightarrow$ ReLU)$\times$3\\
    \cline{2-3}
     & $E_{sq}$ & BiLSTM(512, 2)$\rightarrow$ Average Pooling $\rightarrow$ (Dense(64)$\Rightarrow$ mean, Dense(64)$\Rightarrow$ std)\\
    \cline{2-3}
     & $E_{sp}$ & Identity Mapping\\
    \cline{2-3}
     & $E_{cq}$ & BiLSTM(512, 2)$\rightarrow$RNN(512, 1)$\rightarrow$ (Dense(64)$\Rightarrow$ mean, Dense(64)$\Rightarrow$ std)\\
    \cline{2-3}
     & $E_{cp}$ & BiLSTM(512, 2)$\rightarrow$ (Dense(64)$\Rightarrow$ mean, Dense(64)$\Rightarrow$ std)($\rightarrow$ Linear Classifier)\\
    \cline{2-3}
     & \multirow{2}{*}{$D_{pre}$} & (InstanceNorm2D$\rightarrow$ Conv1D(512, 5, 2, 1)$\rightarrow$ ReLU)$\times$3\\
    \cline{3-3}
     & &LSTM(512, 1) $\rightarrow$ LSTM(1024, 2) $\rightarrow$ Dense(80)\\    
     \cline{2-3}
     & $D_{post}$ & (Conv1D(512, 5, 2, 1)$\rightarrow$  tanh$\rightarrow$ InstanceNorm2D)$\times$4\\
    \hline
    \hline
    Duration Predictor& \multicolumn{2}{|c|}{nn.Embedding$\rightarrow$MHA(256, 128, 128, 2)$\rightarrow$ Conv1D(256, 3, 2, 1) $\rightarrow$ Dense(1)}\\
    \hline
    \hline
    FA2UA & \multicolumn{2}{|c|}{nn.Embedding$\rightarrow$BiLSTM(256, 3)$\rightarrow$ Linear Classifier} \\
    \hline
    \end{tabular}
\caption{The UTTS network architecture in detail. For Conv1D, the configuration is (filter size, kernel size, padding, stride). For Multi-Head-Attention (MHA), the configuration is (model dimension, key dimension, value dimension, $\#$heads). For LSTM/BiLTSM/RNN, the configuration is (hidden dim, layers). For Dense layer, the configuration is (output dim). }
\label{UTTS-model-table}
\end{table*}

\subsubsection{Training Experiments of each components}
For C-DSVAE system, we first train it without content condition (i.e. DSVAE \cite{DSVAE-VC}). Next, we explore the C-DSVAE training with different content conditions, training strategies and datasets, as noted in the bracket. For instance,
C-DSVAE(UA)/C-DSVAE(FA) indicates that the training is conditioned on UA/FA, and C-DSVAE(UA/FA+MP) represents the mask prediction training along with either C-DSVAE(UA) or C-DSVAE(FA). By default C-DSVAE is trained on VCTK corpus unless otherwise specified. For example, C-DSVAE(UA/FA+MP+LibriTTS) shows that C-DSVAE is pretrained on LibriTTS inherited with the C-DSVAE(UA/FA+MP) setting and is then fine-tuned on VCTK under the same configuration. 

In our experiments, the speaker-aware duration predictor and the FA2UA module are trained with LibriTTS for two major reasons: 1) LibriTTS serves as a big external dataset introducing more speaker and phonetic varibilities to the UTTS AMs; 2) we justify that using out of domain speech data (instead of paired VCTK speech and textual data), UTTS can build robust text-to-speech transformations. And the following experimental results demonstrate more details about the systems we developed.

\subsubsection{UTTS Inference}

We use the real-time factor (RTF) to denote the model efficiency in TTS inference. More training hyperparameters can be found in Section \ref{training_config}. The training and inference latency tests are conducted on a server with 48 Intel Xeon CPUs and 4 NVIDIA V100 GPU. We set the batchsize to 1 for inference. The RTF for UTTS is 2.7e-2. Our model can be viewed as an unsupervised AM version of Fastspeech 2, with an additional module that maps text to unsupervised alignment as the input to UTTS AM.   

\subsection{Evaluations on C-DSVAEs}

\subsubsection{C-DSVAE content embeddings and speaker embeddings}
\begin{figure*}[htbp]
    \begin{minipage}[b]{0.18\linewidth}
        \centering
        \centerline{\includegraphics[height=3.5cm, trim=0.8cm 0.5cm 0.8cm 0.5cm,width=4cm]{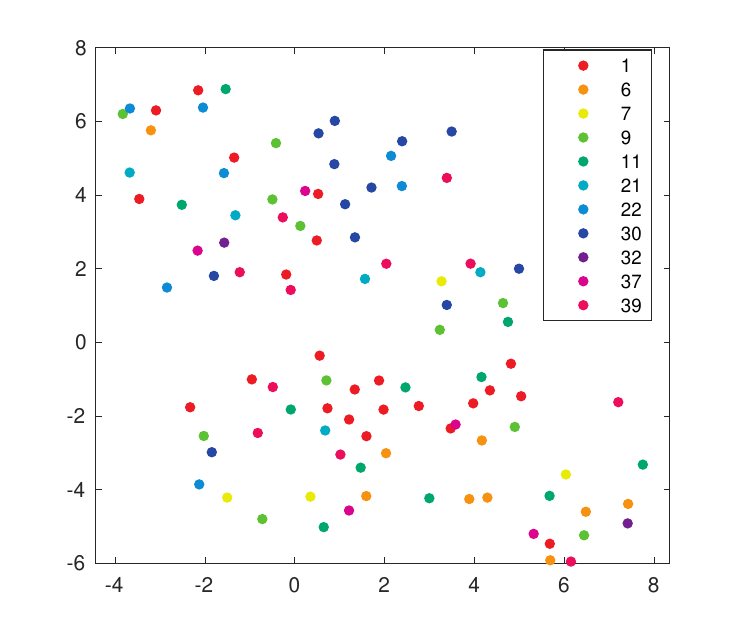}}
        \centerline{\small(a) DSVAE }\medskip
      \end{minipage}
      \hfill
      \begin{minipage}[b]{0.18\linewidth}
        \centering
        \centerline{\includegraphics[height=3.5cm, trim=0.8cm 0.5cm 0.8cm 0.5cm,width=4cm]{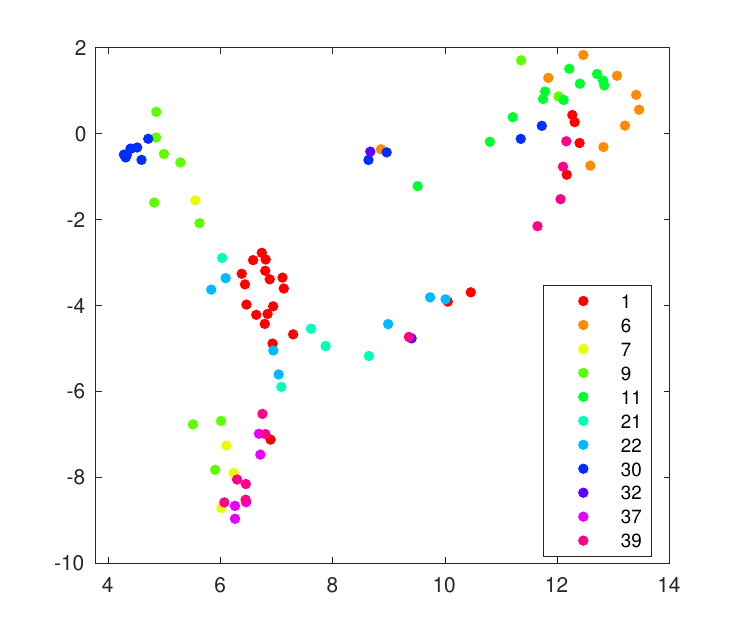}}
        \centerline{\small(b) C-DSVAE(UA) }\medskip
      \end{minipage}
            \hfill
    \begin{minipage}[b]{0.18\linewidth}
      \centering
      \centerline{\includegraphics[height=3.5cm,trim=0.8cm 0.5cm 0.8cm 0.5cm, width=4cm]{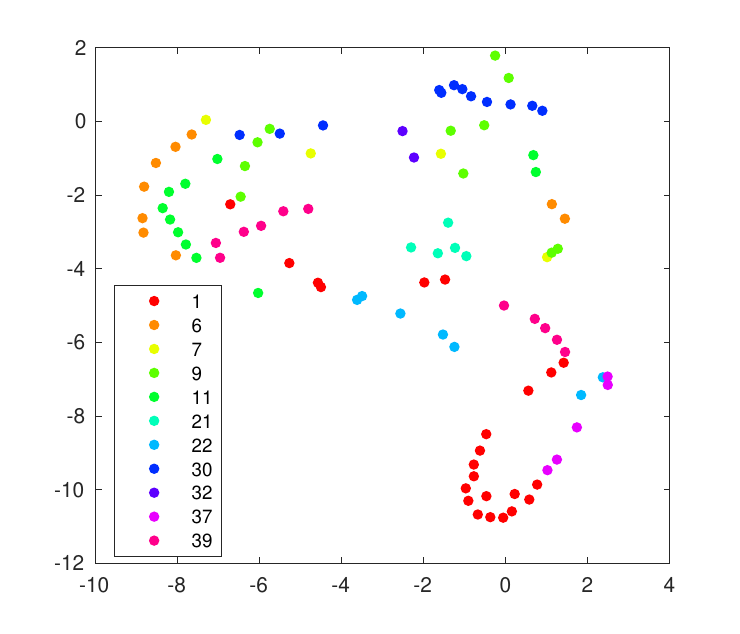}}
      \centerline{(c) C-DSVAE(UA+MP) }\medskip
    \end{minipage}
      \hfill
      \begin{minipage}[b]{0.18\linewidth}
        \centering
        \centerline{\includegraphics[height=3.5cm,trim=0.8cm 0.5cm 0.8cm 0.5cm, width=4cm]{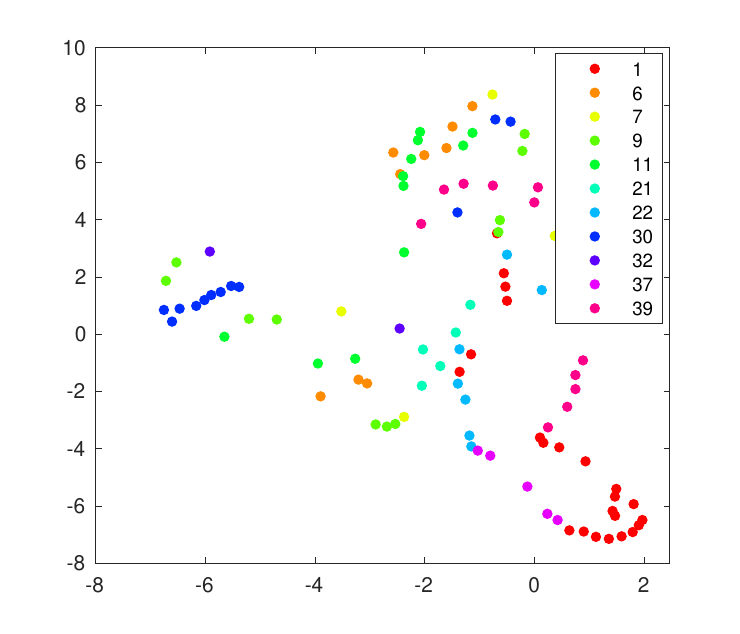}}
        \centerline{(d) C-DSVAE(UA+MP+LibriTTS)}\medskip
      \end{minipage}

    \vspace{-2ex}
    \caption{Visualizations of learned content embeddings.}
    \label{fig:tsne}
\end{figure*}

We evaluate the disentanglement performance of C-DSVAE system via both the t-SNE~\cite{tsne} analysis and objective analysis on content and speaker embeddings. 
\paragraph{t-SNE Analysis}Given pre-trained models, the same utterance that is randomly selected from VCTK test set is taken as input and the t-SNE visualization of the content embeddings is made, where the labels are the forced alignment. We assume that better phonetic discriminativity indicates better disentanglement. As shown in Fig.~\ref{fig:tsne}, C-DSVAE(UA) outperforms DSVAE significantly in terms of phoneme discriminativity, which was also reported in~\cite{C-DSVAE-lian}. It is reasonable that C-DSVAE(UA+MP) and C-DSVAE(UA+MP+LibriTTS) delivers better phonetic structure than C-DSVAE(UA) because masked prediction training does help encode the contextualized and structured phonetic information into the content embeddings. It is also observable that C-DSVAE(UA+MP) delivers better inter-class but worse intra-class phoneme discriminativity, which is assumed to be caused by involving out-of-domain LibriTTS corpus during the training.

\paragraph{Objective Analysis} We perform speaker verification (SV) on both speaker and content embeddings given pre-trained C-DSVAEs. The experiments are conducted on the VCTK SV trials mentioned in Sec.\ref{datasets}, equal error rate (EER) is the metric. Typically, lower EER on speaker embeddings and higher EER on content embeddings indicate better disentanglement. Additionally, the pre-trained C-DSVAE model is fine-tuned with forced alignment on VCTK training set and the phoneme classification accuracy (Top1) for test set is reported. As observed in Table.\ref{tab:phnclass}, C-DSVAEs outperform DSVAE consistently in all metrics, which is consistent with ~\cite{C-DSVAE-lian}. The C-DSVAE with FA condition outperforms that with UA condition, which is reasonable because FA is expected to capture better phonetic structure then unsupervised labels. However, such difference is small and UA-based C-DSVAE is capable of encoding disentangled speaker and content representation which is also semantically meaningful. This provides the prerequisite for the success of the UTTS systems.

\vspace{-1ex}
\begin{table}[!htbp] 
    \centering
     \resizebox{8cm}{!}{
     
    \begin{tabular}{|c |c|c|c|} 
     \hline
      & \multicolumn{2}{c |}{\textbf{content embd}} &  {\textbf{spk embd}} \\
     model & Phn ACC (\%) & SV EER (\%) & SV EER (\%) \\ [0.5ex] 
     \hline\hline
     DSVAE & 30.5 & 40.5 & 4.8 \\ 
     \hline
     C-DSVAE(UA)& 53.6 & 43.2 & 3.9  \\
     C-DSVAE(UA+MP) & 68.2 & 41.6 & 4.1 \\
     C-DSVAE(UA+MP+LibriTTS) & 72.1 & 44.5 & 3.6 \\
     \hline
     C-DSVAE(FA) & 54.6 & 43.4 & 3.8 \\ 
     C-DSVAE(FA+MP) & 70.5 & 42.3 & 3.6 \\
     C-DSVAE(FA+MP+LibriTTS) & 73.3 & 45.6 & 3.3 \\ 
     \hline
    \end{tabular}}
    \caption{Objective analysis of content embeddings and speaker embeddings.}
    \label{tab:phnclass}
\end{table}
\subsubsection{Voice conversion}
The other way to evaluate the disentanglement of C-DSVAE system is to perform voice conversion, as mentioned in Sec.~\ref{voice-conversion-overview}. 
\begin{table*}[th]
     \scriptsize
     \centering
     \resizebox{10cm}{!}{
     \begin{tabular}{|c||c|c||c|c|}
     \hline
   & \multicolumn{2}{c ||}{\bf{seen to seen}} & \multicolumn{2}{c |}{\bf{unseen to unseen}}  \\
    model & naturalness & similarity  & naturalness & similarity\\
     \hline 
     \hline
         
    DSVAE(WaveNet)~\cite{DSVAE-VC} & 3.38$\pm$0.08  & 3.46$\pm$0.07 & 3.36$\pm$0.10  & 3.56$\pm$0.08   \\
   DSVAE(HiFi-GAN)~\cite{C-DSVAE-lian} & 3.56$\pm$0.07  & 3.60$\pm$0.08 & 3.52$\pm$0.08  & 3.62$\pm$0.11   \\
   \hline
   \hline
    C-DSVAE(UA) & 3.78$\pm$0.07  & 3.86$\pm$0.06 & 3.75$\pm$0.07  & 3.82$\pm$0.10   \\
    C-DSVAE(UA+MP) & \textbf{3.85$\pm$0.06}  & \textbf{3.92$\pm$0.08} & \textbf{3.86$\pm$0.06}  & \textbf{ 3.88$\pm$0.09 }   \\
    C-DSVAE(UA+MP+LibriTTS) & 3.80$\pm$0.07  & 3.78$\pm$0.09 & 3.82$\pm$0.09  & 3.86$\pm$0.08   \\
  
     \hline 
     \end{tabular}
      }
       \caption{ The MOS (95\% CI) test on voice conversion samples with different C-DSVAE models.}
     \label{mos} 

\end{table*}
We conduct a MOS test to evaluate the proposed C-DSVAE models for zero-shot non-parallel voice conversion. The evaluation setup is similar to ~\cite{C-DSVAE-lian} and we split the evaluation into to two sets. The first set is a collection of seen speakers sampled from training set and the second set is a collection of unseen speakers sampled from test set. Details on train/test splition of VCTK is introduced in~\ref{datasets}. We randomly draw 30 pairs from each set, with 10 pairs for male to male conversion, 10 pairs for female to female conversion and 10 pairs for cross-gender conversion. The listener needs to give a score for each sample in a test case according to the criterion: 1 = Bad; 2 = Poor; 3 = Fair; 4 = Good; 5 = Excellent. The final score for each model is calculated by averaging the collected results from 18 subjects. As introduced in the paper, we employ HiFi-GAN~\cite{hifigan} as the neural vocoder to convert the mel spectrogram to the waveform. We utilized the samples from two DSVAE based systems for a fair comparison~\cite{C-DSVAE-lian}. Table \ref{mos} shows the MOS results of different models. Since we have different subjects for MOS test and there is always inevitable randomness in subjective evaluation. That is why we observe a difference in C-DSVAE performance between Table \ref{mos} and~\cite{C-DSVAE-lian}. However, the trends that revealed from Table \ref{mos} are still in line with~\cite{C-DSVAE-lian}.

 As illustrated in the Table \ref{mos},  comparing with the WaveNet, HiFi-GAN produces consistently better MOS scores given the same acoustic features. C-DSVAE with unsupervised alignment (UA) achieves 0.24 performance boost in contrast with the best DSVAE voice conversion system. Of all the systems, the C-DSVAE(UA+MP) produced the best objective scores in terms of naturalness and similarity under both seen to seen and unseen to unseen scenarios. And C-DSVAE(UA+MP+LibriTTS) with additional LibriTTS training data does not outperform the VCTK only C-DSVAE(UA+MP) system, which indicates that the domain mismatch exists in the self-supervised representation learning frameworks. It is noted that, the developed voice conversion systems are not sensitive to seen and unseen speech utterances, which is a good attribute from system generalization perspective. However, the relatively larger variance values in the unseen condition shows that unseen cases are still more challenging than the seen cases. 

\subsection{Evaluations on UTTSs}
\subsubsection{MOS test}
We conduct MOS tests in terms of naturalness and speaker similarity on VCTK speakers. 18 subjects are asked to make judgement about the synthesized speech samples. Since our UTTS system is a zero-shot voice cloning format, we also ask the subjects to evaluate the similarity between the UTTS voice and the target voice. We use a standard 5-scale rate to label the speech. 

\begin{table*}[!htbp] 
    \centering
     \resizebox{15.5cm}{!}{
     
    \begin{tabular}{|c |c|c|c|c||c|c|c|c|} 
     \hline
      & \multicolumn{2}{c |}{\textbf{seen target}} & \multicolumn{2}{c ||} {\textbf{unseen target}} &  \multicolumn{2}{c |}{\textbf{male sample}} & \multicolumn{2}{c |} {\textbf{female sample}} \\
     UTTS model & naturalness & similarity & naturalness & similarity & naturalness & similarity & naturalness & similarity  \\ 
     \hline\hline
     GT & \multicolumn{4}{c ||} {4.45} & \multicolumn{2}{c |} {4.52} & \multicolumn{2}{c |} {4.39} \\ 
     \hline
     C-DSVAE(UA)& 3.22 & 3.56 & 3.28 & 3.51 & 3.38 & 3.64 & 3.26 & 3.44 \\
     C-DSVAE(UA+MP) & 3.52 & 3.72 & 3.56 & 3.64 & 3.65 & 3.76 & 3.45 & 3.62 \\
     C-DSVAE(UA+MP+Dual) & \textbf{3.76} & \textbf{3.82} & \textbf{3.72} & \textbf{3.74} & \textbf{3.82} & \textbf{3.90} & \textbf{3.67} & \textbf{3.71} \\
     C-DSVAE(UA+MP+Dual+LibriTTS) & 3.58 & 3.63 & 3.61 & 3.70 & 3.68 & 3.73 & 3.56 & 3.60 \\
     \hline
    \end{tabular}}
    \caption{A multi-speaker UTTS MOS test of naturalness and similarity on VCTK. }
    \label{mos1}
\end{table*}
As shown from Table \ref{mos1}, compared with the vanilla C-DSVAE(UA) AM, the masked unit prediction multi-task training in the content prior model of C-DSVAE does helps the overall speech quality. Having LibriTTS for pretraining helps but not so significantly. If the dual prior encoder training is applied to the C-DSVAE(UA+MP) system, both naturalness and similarity are improved clearly, showing the importance of prior modeling in this VAE based non-autoregressive TTS model. At the same time, the gap between synthesised speech and the ground truth in naturalness indicates that some form of prosody are currently missing in the proposed UTTS systems.  
\subsubsection{Speech intelligibility} \label{asr}
Results of intelligibility evaluation using a Librispeech ASR model\footnote{\url{https://huggingface.co/espnet/simpleoier\_librispeech\_asr\_train\_asr\_conformer7\_hubert\_ll60k\_large\_raw\_en\_bpe5000\_sp}} are also listed in Table \ref{asr_gender}. We generated 1600 utterances each model for this portion of study. Basically, the objective and subjective evaluation show a similar trend. Speech from the best UTTS model (C-DSVAE(UA+MP+Dual)) achieves a 15.1\% WER, the gap between ground truth and UTTS speech suggests that the proposed method can be further improved by either improving the FA2UA mapping module or the content prior modeling. In Table \ref{asr_gender}, we also list the ASR based speech intelligibility test results on UTTS samples by gender. As shown from the table, the ASR system does not show much performance preference w.r.t. different genders on real speech. However, the ASR results for male and female synthesised samples are showing a big difference. The much worse CER and WER for female generated speech across all 4 UTTS systems indicates that generating female voice seems to be more difficult in the multi-speaker TTS systems. This is also observed in Table \ref{mos1}, where gender dependent MOS scores indicates that the current systems are generating better male voices.

\begin{table}[!htbp] 
    \centering
     \resizebox{8.6cm}{!}{
     
    \begin{tabular}{|c|c|c||c|c|c|c|} 
     \hline
       & \multicolumn{2}{c ||} {\textbf{overall}} & \multicolumn{2}{c |}{\textbf{male}} &  \multicolumn{2}{c |}{\textbf{female}} \\
     UTTS model &  CER (\%) & WER (\%) & CER (\%) & WER (\%) & CER (\%) & WER (\%) \\ 
     \hline\hline
     GT  & 2.3 & 7.0& 2.4 & 7.3 & 2.2 & 6.7 \\ 
     \hline
     C-DSVAE(UA) & 11.3 & 19.3 & 8.8 & 16.9 & 13.8 & 21.6  \\
     C-DSVAE(UA+MP)  & 9.7 & 16.5 & 6.1 & 12.8 & 13.3 & 20.2 \\
     C-DSVAE(UA+MP+Dual)  & \textbf{8.3} & \textbf{15.1} & \textbf{5.2} & \textbf{11.5} & \textbf{11.4} & \textbf{18.7} \\
     C-DSVAE(UA+MP+Dual+LibriTTS)  & 10.8 & 18.2 & 7.3  & 14.7 & 14.5& 21.8\\
     \hline
    \end{tabular}}
    \caption{An ASR intelligibility test on gender dependent UTTS samples. }
    \label{asr_gender}
\end{table}

\subsubsection{Speaker similarity between the real and UTTS speech}
\label{spk_similarity}

\begin{figure}[ht]
    \centering
    \includegraphics[height=4.5cm,trim=0.8cm 0.8cm 0.8cm 0.8cm, width=6.cm]{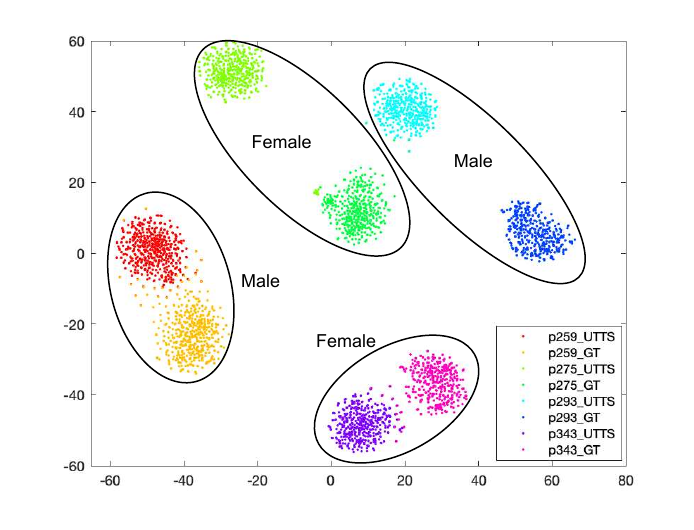}
    \caption{T-SNE visualization of utterances of 4 VCTK speakers. }
    \label{fig:sv_similarity}
\end{figure}

Since UTTS is a zero-shot voice cloning TTS system, Table \ref{mos} demonstrates how human evaluates the similarity between the synthesised and target speech. In the meantime, we are interested to know how machine perceives the similarity and difference. For this purpose, we use a pretrained speaker embedding system to extract speaker embeddings from real and synthesised utterances, and project 
them to a 2D T-SNE~\cite{zhang2018text,tsne} space. Fig.\ref{fig:sv_similarity} illustrates the speaker embedding projections of 4 speakers (P259, P275, P293 and P343). P259 and P293 are used for the UTTS AM training, while P275 and P343 are never used in the training stage. 

From the clusters in Fig.\ref{fig:sv_similarity}, it seems that each generated speaker localizes at its unique speaker space. The similarity between real and UTTS speech is showing some randomness. For example, P259 (male \& seen) and P343 (female \& unseen) are showing much obvious speaker similarities in speaker embedding space, while P275 (female \& unseen) and P293 (male \& seen) demonstrates a more separated speaker attribute in the same space. All the synthesised utterances are formulating a distinct distribution that is different with the target speaker distribution, which indicates that the synthetic speech from the current UTTS systems still can be detected with a normal speaker verification system.

\section{Conclusion}
\label{conclusion}
In this study, we proposed a novel method to develop TTS systems without using the paired TTS training data. Specifically, we explored C-DSVAE, a self-supervised model that learned disentangled intermediate content and speaker representations with unsupervised alignment as the prior condition, and generated high quality speech with the latent speaker and content features using a decoder network. We employed C-DSVAE as the backbone module for developing the UTTS systems. For that purpose, we proposed a simple yet effective forced alignment to unsupervised alignment module to bridge the TTS front-end and the C-DSVAE. To further improve the speech quality and naturalness, we proposed to have a dual prior encoder reconstruction training strategy in our VAE based AM training, which significantly alleviated the mismatch between training and test phase of UTTS inference. Preliminary experiments demonstrated that our proposed UTTS systems can produce speech with an impressive quality. The newly developed method shows great potential for its self-supervised/unsupervised nature that can utilize general quality speech data for future needs.

At the same time, we found the UTTS generated samples still have some limitations in speech quality, speech expressiveness and speaker similarity. The the speech fidelity related issues can be addressed by introducing more advanced text encoding networks or methods~\cite{vitsl,ho2020denoising}. While the latter one can be improved with more data involved in the self-supervised or semi-supervised speaker representation training~\cite{zhang2022c3}. Thus, our study serves as the preliminary research work, proving the concept of unsupervised TTS AM, while leaving observed inadequacies as the future work.

\bibliographystyle{IEEEtran}
\bibliography{refs}
\section{Biography Section}

\begin{IEEEbiography}[{\includegraphics[width=1in,height=1.25in,clip,keepaspectratio]{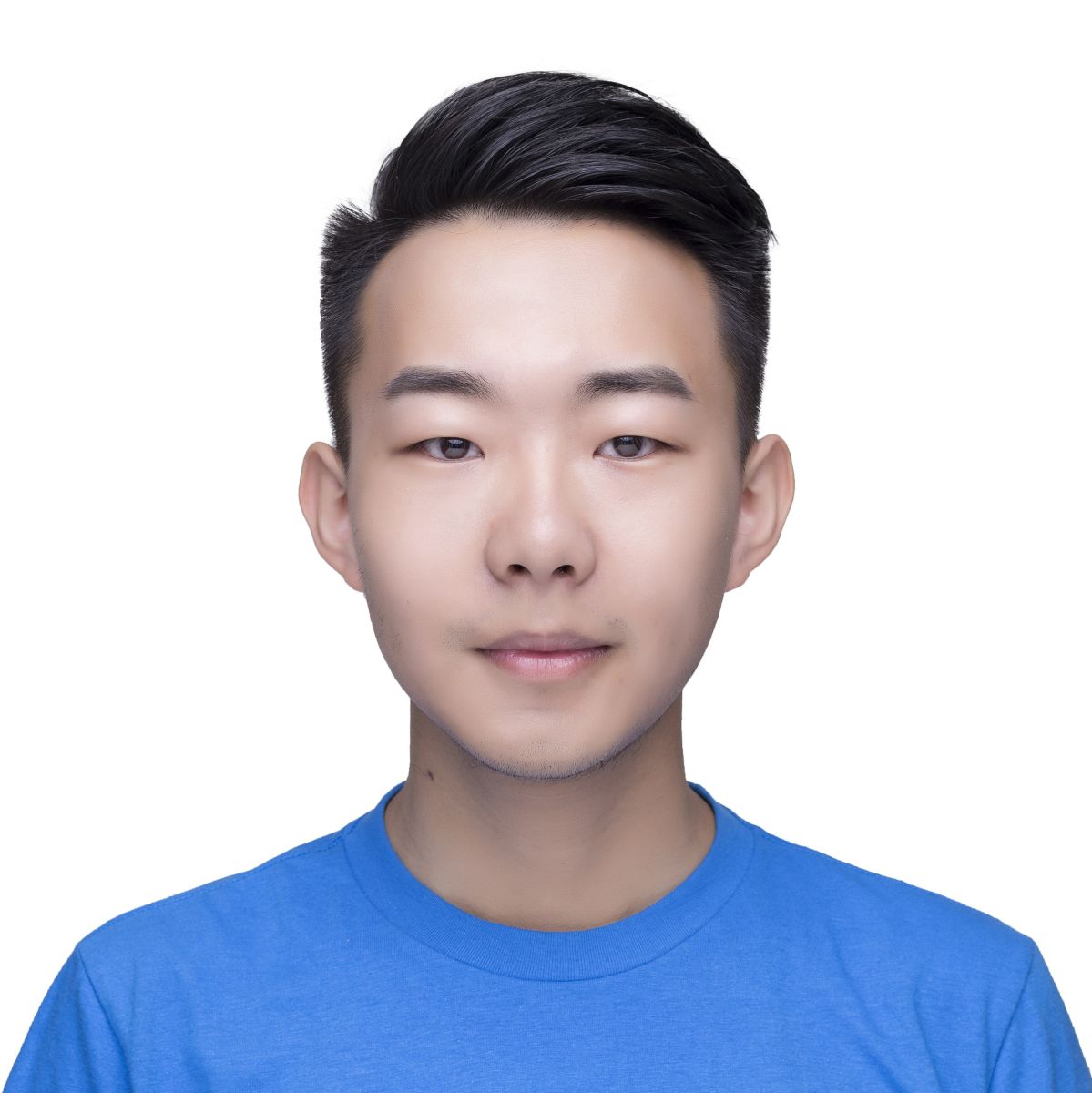}}]{Jiachen Lian} (M'22) received the B.S. degree in electrical engineering from Zhejiang University, Hangzhou, China and the M.S. degree in electrical and computer engineering from Carnegie Mellon University, USA, in 2019 and 2021, respectively. He is currently pursuing his Ph.D. degree in EECS from UC Berkeley, USA.  His research interests span deep learning and speech processing. 
\end{IEEEbiography}

\begin{IEEEbiography}[{\includegraphics[width=1in,height=1.25in,clip,keepaspectratio]{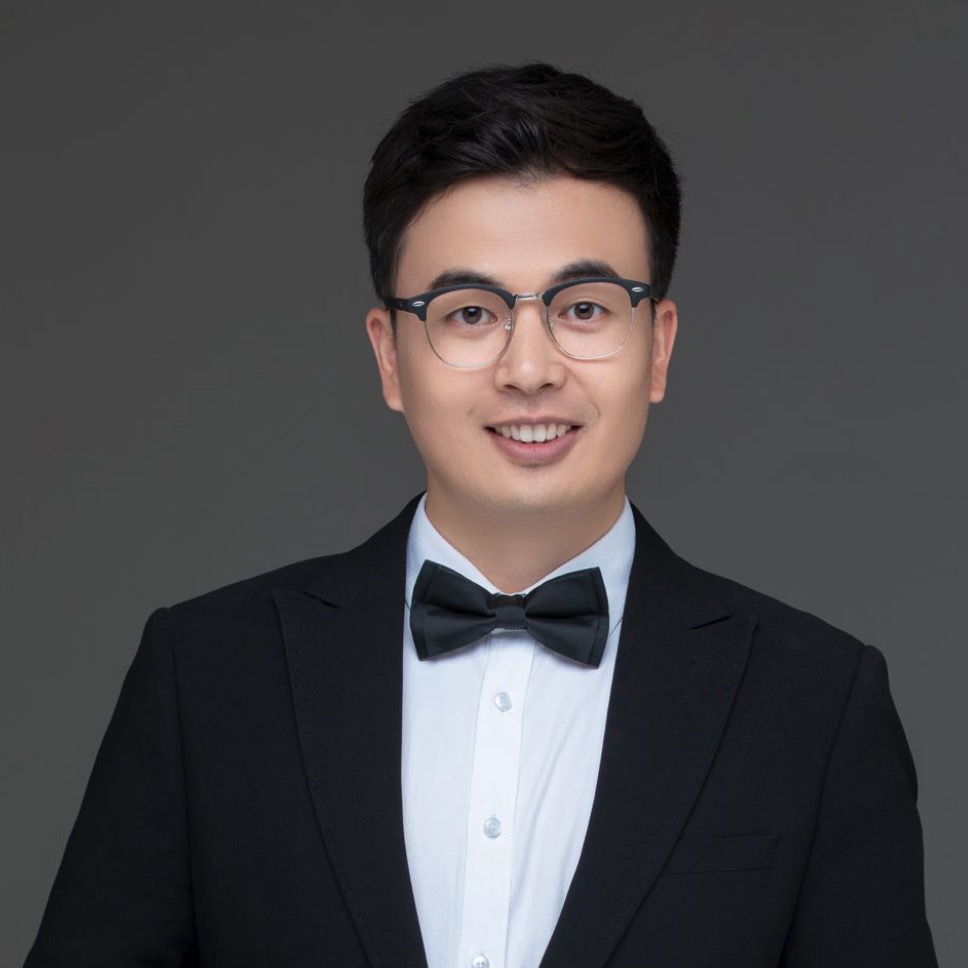}}]{Chunlei Zhang} (M'16) received the B.S. degree in environmental engineering and the M.S. degree in acoustics from Northwestern Polytechnical University, Xi’an, China, in 2011 and 2014, respectively. He received his Ph.D. degree in electrical and computer engineering from The University of Texas at Dallas, Richardson, TX, USA, in 2018. He joined the Tencent AI Lab, Bellevue, WA, USA, in 2019, where he is currently a Senior Research Scientist. His research interests include automatic speech recognition, speaker recognition/diarization and text-to-speech synthesis. He has been a member of the ISCA and IEEE since 2016. 
\end{IEEEbiography}

\begin{IEEEbiography}[{\includegraphics[width=1in,height=1.25in,clip,keepaspectratio]{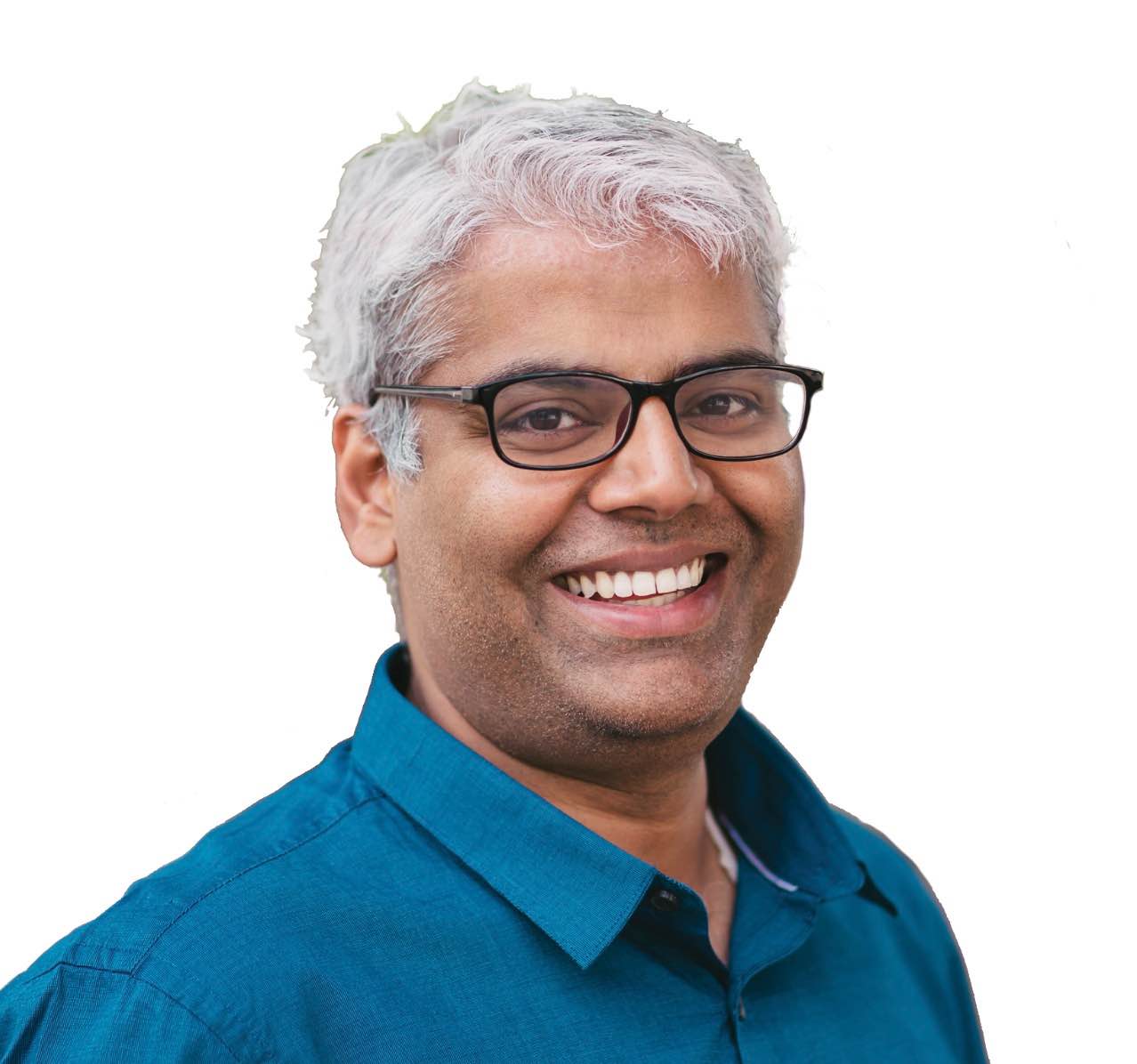}}]{Gopala K. Anumanchipalli} (M'08) received the B.T. and M.S. degree in computer science from International Institute of Information Technology, Hyderabad, India, in 2008. He received his Ph.D. degree in Language and Information Technologies from Carnegie Mellon University, USA, and in electrical and computer engineering from IST, Lisbon, in 2013. He joined the faculty of the Department of Electrical Engineering and Computer Sciences at UC Berkeley in Spring 2021 and continues to hold an adjunct position at Dept. of Neurosurgery at UC San Francisco. He works at the intersection of Speech Processing, Neuroscience, and Artificial Intelligence with an emphasis on human-centered speech and Assistive technologies, including new paradigms for bio-inspired spoken language technologies, automated methods for early diagnosis, characterizing and rehabilitating disordered speech. 
\end{IEEEbiography}

\begin{IEEEbiography}[{\includegraphics[width=1in,height=1.25in,clip,keepaspectratio]{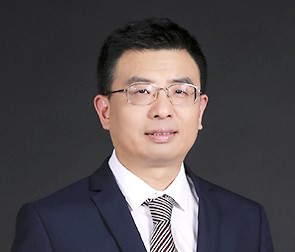}}]{Dong Yu} (M'97 SM'06 F'18) Dr. Dong Yu is an ACM/IEEE/ISCA Fellow. He works at Tencent AI Lab as a distinguished scientist and vice general manager. Prior to joining Tencent in 2017, he worked as a principal researcher at Microsoft Research (Redmond), where he had been since 1998. Dr. Dong Yu’s research focuses on speech recognition and processing and natural language processing. He has published two monographs and over 300 papers. His works have been widely cited and recognized by the prestigious IEEE Signal Processing Society best transaction paper award in 2013, 2016, 2020, and 2022, the 2021 NAACL best long paper award, the 2022 IEEE Signal Processing Magazine best paper award, and the 2022 IEEE Signal Processing Magazine best column award. 

Dr. Dong Yu was the chair of the IEEE Speech and Language Processing Technical Committee in 2021-2022. He has served on the editorial boards of numerous journals and magazines, as well as on the organizing and technical committees of various conferences and workshops. 
\end{IEEEbiography}

\end{document}